\documentclass{kluwer}    

\usepackage{graphicx}

\begin{document}                                                                                   
\begin{article}
\begin{opening}         
\title{Modelling the Extinction Properties of Galaxies} 
\author{G.L. \surname{Granato}} 
\author{L. \surname{Silva}}, 
\author{A. \surname{Bressan}}, 
\institute{Osservatorio Astronomico di Padova, Vicolo
dell'Osservatorio, 5, I-35122 Padova, Italy}
\author{C.G. Lacey},
\institute{SISSA, Via Beirut 2-4, I-34014 Trieste, Italy}
\author{C.M. \surname{Baugh}}
\author{S. \surname{Cole}}
\author{C.S. \surname{Frenk}}
\institute{Physics Department, Durham University, South Road,
Durham DH1 3LE, UK}

\runningauthor{G.L. Granato et al.}
\runningtitle{Modelling the Extinction Properties of Galaxies}


\begin{abstract} Recently (Granato, Lacey, Silva et al. 2000, astro-ph/0001308)
we have combined our spectrophotometric galaxy evolution code which includes
dust reprocessing (GRASIL, Silva et al. 1998) with semi-analytical galaxy
formation models (GALFORM, Cole et al. 1999). One of the most characteristic
features of the former is that the dust is divided in two main phases:
molecular cloud complexes, where stars are assumed to be born, and the diffuse
interstellar medium. As a consequence, stellar populations of different ages
have different geometrical relationships with the two phases, which is
essential in understanding several observed properties of galaxies, in
particular those undergoing major episodes of star formation at any redshift.
Indeed, our merged GRASIL+GALFORM model reproduces fairly well the SEDs of
normal spirals and starbursts from the far-UV to the sub-mm and their internal
extinction properties. In particular in the model the observed starburst
attenuation law (Calzetti 1999) is accounted for as an effect of geometry of
stars and dust, and has nothing to do with the optical properties of dust
grains.

\end{abstract}

\end{opening}           

\section{Introduction}  

Semi--analytical models are the key technique to predict galaxy properties in
the framework of hierarchical structure formation. Simplified analytical
descriptions of gas cooling and collapse, star formation, supernovae feedback
and galaxy merging are applied to a Monte Carlo description of the formation
and merging of DM halos. However, semi-analytical models have so far ignored
or treated poorly dust reprocessing.

To cope with this point, which according to several pieces of evidence appears
to be crucial to understand high-z observations, we combined the
semi--analytical galaxy formation model of Cole et al (1999, GALFORM) with the
stellar population + dust model of Silva et al (1998, GRASIL). Both models are
state--of--the--art.

We refer the reader to Granato et al. (2000) for the details. Here we remind
only the basic features of our modelling. GALFORM includes: (1) formation of DM
halos through merging; (2) cooling and collapse of gas in halos to form disks;
(3) star formation in disk regulated by supernovae feedback; (4) merging of
disk galaxies to forms ellipticals and bulges; (5) bursts of star formation
triggered by these mergers; (6) predictions of the radii of disks and
spheroids; (7) star formation and chemical enrichment histories of stars and
gas.

GRASIL (http://grana.pd.astro.it, Silva, Granato, Bressan \& Danese 1998)
includes: (1) a realistic 3D geometry (disk + bulge) with a two phase ISM
(cirrus + Molecular Clouds MCs); (2) birth and early evolution of stars in MCs;
(3) clumpiness of both ISM and stars spatial distributions, with age
dependence; (4) radiative transfer whenever required; (5) dust grain model
including PAHs and quantum heating of small grains, calibrated to fit the MW
extinction law; (6) self consistent computation of thermal status of grains in
each point; (7) effects of AGB dusty envelopes.

The purpose of our first paper is to study the effects of including dust in a
fixed galaxy formation model, chosen previously by Cole et al (1999) to fit the
properties of local galaxies in the optical-NIR. GALFORM provides the star
formation and chemical enrichment histories, the gas mass and various
geometrical parameters of mock catalogs of galaxies at various redshifts.
GRASIL uses these information to predict synthetic SEDs. In this way, now
semi-analytical models can be effectively compared with IR and sub-mm data,
essential to understand the high-z SF history.

We test our models against the observed spectro-photometric properties of
galaxies in the local Universe, assuming a CDM cosmology with $\Omega_0=0.3$
and $\Lambda_0=0.7$. In this contribution we focus our attention on one
particular result, namely our interpretation of the observed starburst
attenuation law. Before doing this, we summarize very briefly our other
findings.

The models reproduce fairly well the SEDs of normal spirals and starbursts from
the far-UV to the sub-mm, and their internal extinction properties. The
starbursts follow the observed relationship between the FIR to UV luminosity
ratio and the slope of the UV continuum.  We compute galaxy luminosity
functions over a wide range of wavelengths, which turn out to be in good
agreement with observational data in the UV (2000\AA), in the B and K bands,
and in the IR ($12-100\mu$m). Finally, we investigate the reliability of some
star formation indicators which are based on the properties of the continuum
SEDs of galaxies. The UV continuum turns out to be a poor star formation
indicator for our models, whilst the infrared luminosity is much more reliable.

\section{Interpretation of the observed starburst attenuation law}

An important problem in the study of star-forming galaxies is to determine the
amount of attenuation of starlight by dust, especially in the UV. This bears
directly on the determination of star formation rates in galaxies from their UV
luminosities.  The differences found between the shapes of the extinction
curves of the Galaxy, the LMC and the SMC below $\lambda\leq$2600\AA\ are often
ascribed to the different metallicities in these systems.

From the optical and UV spectra of a sample of UV-bright starbursts, Calzetti
et al.\ (1994) derived an average {\em attenuation law} characterized by a
shallower far-UV slope than that of the Milky Way extinction law, and by the
absence of the 2175 \AA\ feature. This is at first sight quite surprising,
because the metallicities of these galaxies are mostly similar to that of the
Milky Way, and so they might be expected to have similar dust properties. The
question is then to what degree the differences between the starburst
attenuation law and the Milky Way extinction law are due to the geometry of the
stars and dust, and to what degree they can only be explained by differences in
dust properties.

\begin{figure}[!htb] 
\centerline{\includegraphics[width=7.0truecm]{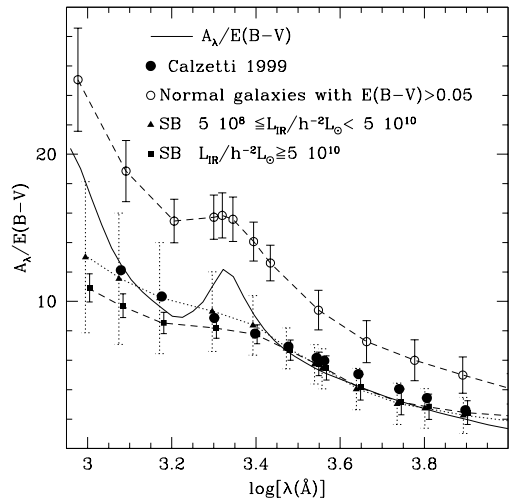}}
\caption{The average dust attenuation curves for starlight in
different classes of galaxies (normal and starburst, SB) in the model
compared with the average Milky Way extinction law (solid line) and
with the Calzetti ``attenuation law'' (filled circles,
with $R'_V=4.05$).  
The error bars show the dispersion of the models around the
mean attenuation curve.}
\label{fig:calz}
\end{figure}

Figure~\ref{fig:calz} compares the average attenuation curves for galaxies from
our model with the empirical ``attenuation law'' obtained for starbursts by
Calzetti (1999). As already remarked, the dust properties we adopt imply an
extinction law characterized by a distinct 2175 \AA\ feature produced by
graphite grains, and well matching the average Milky Way extinction curve. The
model extinction law (solid line in Fig.~\ref{fig:calz}) is the attenuation law
that would be measured if all the dust were in a foreground screen in front of
the stars and no scattered light reached the observer. This geometry is clearly
not realistic as applied to the integrated light from galaxies. In our models,
we have instead a complex and wavelength dependent geometry, where the UV
emitting stars are heavily embedded inside molecular clouds, while the older
stars, mainly emitting in the optical and near infrared, are well mixed with
the diffuse interstellar medium.

All the 3 classes of models in Figure~\ref{fig:calz} show a weak or completely
absent $2175$ \AA\ feature. In particular, the predicted attenuation curve for
the lower luminosity starbursts is remarkably close to the empirical ``Calzetti
law''. This result is an entirely geometrical effect, and did not require us to
assume for starbursts dust properties different from those of the Galaxy, but
rather follows naturally from the assumption that stars are born inside 
optically thick dust clouds and gradually escape.

Indeed, in the far-UV, including the spectral region around the $2175$ \AA\
feature, the global attenuation in the models is strongly contributed, or even
dominated, by the MCs.  The shape of the attenuation curve there has little to
do with the optical properties of grains, because our MCs usually have such
large optical depths that the UV light from stars inside the clouds is
completely absorbed. The wavelength dependence of the attenuation law of the MC
component instead arises from the fact that the fraction of the light produced
by very young stars increases with decreasing wavelength, and at the same time,
the fraction of stars which are inside clouds increases with decreasing age.
The additional attenuation arising in the cirrus component can sometimes
imprint a weak $2175$ \AA\ feature, but this is not the case for the starbursts.

\end{article}
\end{document}